\documentclass[english,twocolumn]{IEEEtran}
\usepackage[T1]{fontenc}
\usepackage[latin9]{inputenc}
\usepackage{verbatim}
\usepackage{amsthm}
\usepackage{amsmath}
\usepackage{amssymb}
\usepackage{esint}

\makeatletter
\theoremstyle{definition}
\newtheorem{assumption}{Assumption}
  \theoremstyle{remark}
  \newtheorem{rem}{\protect\remarkname}
  \theoremstyle{plain}
  \newtheorem{thm}{\protect\theoremname}

\usepackage{cite}
\allowdisplaybreaks
\IEEEoverridecommandlockouts

\makeatother

\usepackage{babel}
\providecommand{\remarkname}{Remark}
\providecommand{\theoremname}{Theorem}

\begin{document}

\title{Model-based reinforcement learning for infinite-horizon approximate
optimal tracking%
\thanks{Rushikesh Kamalapurkar, Lindsey Andrews, Patrick Walters, and Warren
E. Dixon are with the Department of Mechanical and Aerospace Engineering,
University of Florida, Gainesville, FL, USA. Email: \{rkamalapurkar,
landr010, walters8, wdixon\}@ufl.edu.%
}%
\thanks{This research is supported in part by NSF award numbers 1161260 and
1217908, ONR grant number N00014-13-1-0151, and a contract with the
AFRL Mathematical Modeling and Optimization Institute. Any opinions,
findings and conclusions or recommendations expressed in this material
are those of the authors and do not necessarily reflect the views
of the sponsoring agency.%
}%
\thanks{Submitted to the special issue on \textit{New Developments in Neural
Network Structures for Signal Processing, Autonomous Decision, and
Adaptive Control }%
}}

\author{Rushikesh Kamalapurkar, Lindsey Andrews, Patrick Walters, and Warren
E. Dixon}
\maketitle
\begin{abstract}
This paper provides an approximate online adaptive solution to the
infinite-horizon optimal tracking problem for control-affine continuous-time
nonlinear systems with unknown drift dynamics. Model-based reinforcement
learning is used to relax the persistence of excitation condition.
Model-based reinforcement learning is implemented using a concurrent
learning-based system identifier to simulate experience by evaluating
the Bellman error over unexplored areas of the state space. Tracking
of the desired trajectory and convergence of the developed policy
to a neighborhood of the optimal policy are established via Lyapunov-based
stability analysis. Simulation results demonstrate the effectiveness
of the developed technique.\end{abstract}


\section{Introduction}

In the past few decades, reinforcement learning (RL)-based techniques
have been effectively utilized to obtain online approximate solutions
to optimal control problems for systems with finite state-action spaces,
and stationary environments (cf. \cite{Sutton1998,Bertsekas2007}).
However, progress for systems with continuous state-action spaces
has been slow due to various technical challenges (cf. \cite{Mehta.Meyn2009,Deisenroth2010}).
Various implementations of RL-based learning strategies to solve deterministic
optimal regulation problems can be found in results such as \cite{Doya2000,Abu-Khalaf.Lewis.ea2006,Padhi2006,Al-Tamimi2008,Chen2008,Dierks2009,Vamvoudakis2010,Bhasin.Kamalapurkar.ea2013a,Zhang.Liu.ea2013,Liu.Wei2014,Modares.Lewis.ea2014,Yang.Liu.ea2014}.

Offline and online approaches to solve infinite-horizon tracking problems
are proposed in results such as \cite{Zhang2008a,Dierks2009a,Zhang.Cui.ea2011,Wei.Liu2013,Kiumarsi.Lewis.ea2014,Qin.Zhang.ea2014}.
Results such as \cite{Dierks2009a,Kiumarsi.Lewis.ea2014,Qin.Zhang.ea2014,Kamalapurkar.Dinh.ea2015}
solve optimal tracking problems for linear and nonlinear systems online,
where persistence of excitation (PE) of the error states is used to
establish convergence. In general, it is impossible to guarantee PE
a priori; hence, a probing signal designed using trial and error is
added to the controller to ensure PE. However, the probing signal
is not considered in the stability analysis. In this paper, the objective
is to employ data-driven model-based RL to design an online approximate
optimal tracking controller for continuous-time uncertain nonlinear
systems under a relaxed finite excitation condition.

RL in systems with continuous state and action spaces is realized
via value function approximation, where the value function corresponding
to the optimal control problem is approximated using a parametric
universal approximator. The control policy is generally derived from
the approximate value function; hence, obtaining a good approximation
of the value function is critical to the stability of the closed-loop
system. In trajectory tracking problems, the value function depends
explicitly on time. Since universal function approximators can approximate
functions with arbitrary accuracy only on compact domains, value functions
for infinite-horizon optimal tracking problems can not be approximated
with arbitrary accuracy \cite{Zhang2008a,Kamalapurkar.Dinh.ea2015}. 

If the desired trajectory can be expressed as the output of an autonomous
dynamical system, then the value function can be expressed as a stationary
(time-independent) function of the state and the desired trajectory.
Hence, universal function approximators can be employed to approximate
the value function with arbitrary accuracy by using the system state,
augmented with the desired trajectory, as the training input (cf.
\cite{Zhang2008a,Kiumarsi.Lewis.ea2014,Qin.Zhang.ea2014,Kamalapurkar.Dinh.ea2015}).

The technical challenges associated with the nonautonomous nature
of the trajectory tracking problem are addressed in the author's previous
work in \cite{Kamalapurkar.Dinh.ea2015}, where it is established
that under a matching condition on the desired trajectory, the optimal
trajectory tracking problem can be reformulated as a stationary optimal
control problem. Since the value function associated with a stationary
optimal control problem is time-invariant, it can be approximated
using traditional function approximation techniques. 

The aforementioned reformulation in \cite{Kamalapurkar.Dinh.ea2015}
requires computation of the steady-state tracking controller, which
depends on the system model; hence, the development in \cite{Kamalapurkar.Dinh.ea2015}
requires exact model knowledge. Obtaining an accurate estimate of
the desired steady-state controller, and injecting the resulting estimation
error in the stability analysis are the major technical challenges
in extending the work in \cite{Kamalapurkar.Dinh.ea2015} to uncertain
systems. In this paper and in the preliminary work in \cite{Kamalapurkar.Andrews.ea2014},
a concurrent learning (CL)-based system identifier is used to estimate
the desired steady-state controller and model-based RL is used to
simulate experience by evaluating the Bellman error (BE) over unexplored
areas of the state space \cite{Chowdhary2010a,Chowdhary.Johnson2011a,Chowdhary.Yucelen.ea2012,Kamalapurkar.Andrews.ea2014}.
The error between the actual steady-state controller and its estimate
is included in the stability analysis by formulating the Hamilton-Jacobi-Bellman
equation in terms of the actual steady-state controller, and the effectiveness
of the developed technique is demonstrated via numerical simulations. 

The main contributions of this work include: 1) Approximate model
inversion using a CL-based system identifier to approximate the desired
steady-state controller in the presence of uncertainties in the drift
dynamics, 2) Implementation of model-based RL to relax the PE condition
to a finite excitation condition, 3) Simulation results that demonstrate
approximation of the optimal policy without an added exploration signal.

\section{Problem formulation and exact solution\label{sec:Problem-formulation-and}}

Consider a control affine system described by the differential equation
$\dot{x}=f\left(x\right)+g\left(x\right)u,$ where $x\in\mathbb{R}^{n}$
denotes the state, $u\in\mathbb{R}^{m}$ denotes the control input,
and $f:\mathbb{R}^{n}\to\mathbb{R}^{n}$ and $g:\mathbb{R}^{n}\to\mathbb{R}^{n\times m}$
are locally Lipschitz continuous functions that denote the drift dynamics,
and the control effectiveness, respectively.%
\footnote{For notational brevity, unless otherwise specified, the domain of
all the functions is assumed to be $\mathbb{R}_{\geq0}$. Furthermore,
time-dependence is suppressed in equations and definitions. For example,
the trajectory $x:\mathbb{R}_{\geq0}\to\mathbb{R}^{n}$ is defined
by abuse of notation as $x\in\mathbb{R}^{n}$ and unless otherwise
specified, an equation of the form $f+h\left(y,t\right)=g\left(x\right)$
is interpreted as $f\left(t\right)+h\left(y\left(t\right),t\right)=g\left(x\left(t\right)\right)$
for all $t\in\mathbb{R}_{\geq0}$.%
} The control objective is to optimally track a time-varying desired
trajectory $x_{d}\in\mathbb{R}^{n}$. To facilitate the subsequent
control development, an error signal $e\in\mathbb{R}^{n}$ is defined
as $e\triangleq x-x_{d}.$ Since the steady-state control input that
is required for the system to track a desired trajectory is, in general,
not identically zero, an infinite-horizon total-cost optimal control
problem formulated in terms of a quadratic cost function containing
$e$ and $u$ always results in an infinite cost. To address this
issue, an alternative cost function is formulated in terms of the
tracking error and the mismatch between the actual control signal
and the desired steady-state control \cite{Zhang2008a,Kiumarsi.Lewis.ea2014,Qin.Zhang.ea2014,Kamalapurkar.Dinh.ea2015}.
The following assumptions facilitate the determination of the desired
steady-state control.
\begin{assumption}
\label{ass:CLTg}\cite{Kamalapurkar.Dinh.ea2015} The function $g$
is bounded, the matrix $g\left(x\right)$ has full column rank for
all $x\in\mathbb{R}^{n}$, and the function $g^{+}:\mathbb{R}^{n}\rightarrow\mathbb{R}^{m\times n}$
defined as $g^{+}\triangleq\left(g^{T}g\right)^{-1}g^{T}$ is bounded
and locally Lipschitz. 
\end{assumption}

\begin{assumption}
\label{ass:CLTAhd}\cite{Kamalapurkar.Dinh.ea2015} The desired trajectory
is bounded by a known positive constant $d\in\mathbb{R}$ such that
$\left\Vert x_{d}\right\Vert \leq d$, and there exists a locally
Lipschitz function $h_{d}:\mathbb{R}^{n}\rightarrow\mathbb{R}^{n}$
such that $\dot{x}_{d}=h_{d}\left(x_{d}\right)$ and 
\[
g\left(x_{d}\right)g^{+}\left(x_{d}\right)\left(h_{d}\left(x_{d}\right)-f\left(x_{d}\right)\right)=h_{d}\left(x_{d}\right)-f\left(x_{d}\right),
\]
for all $t\in\mathbb{R}_{\geq t_{0}}$.
\end{assumption}
Based on Assumptions \ref{ass:CLTg} and \ref{ass:CLTAhd}, the steady-state
control policy $u_{d}:\mathbb{R}^{n}\to\mathbb{R}^{m}$ required for
the system to track the desired trajectory $x_{d}$ can be expressed
as $u_{d}\left(x_{d}\right)=g_{d}^{+}\left(h_{d}\left(x_{d}\right)-f_{d}\right),$
where $f_{d}\triangleq f\left(x_{d}\right)$ and $g_{d}^{+}\triangleq g^{+}\left(x_{d}\right)$.
The error between the actual control signal and the desired steady-state
control signal is defined as $\mu\triangleq u-u_{d}\left(x_{d}\right).$
Using $\mu$, the system dynamics can be expressed in the autonomous
form 
\begin{equation}
\dot{\zeta}=F\left(\zeta\right)+G\left(\zeta\right)\mu,\label{eq:CLTzetadyn}
\end{equation}
 where the concatenated state $\zeta\in\mathbb{R}^{2n}$ is defined
as $\zeta\triangleq\left[e^{T},\: x_{d}^{T}\right]^{T},$ and the
functions $F:\mathbb{R}^{2n}\to\mathbb{R}^{2n}$ and $G:\mathbb{R}^{2n}\to\mathbb{R}^{2n\times m}$
are defined as 
\[
F\left(\zeta\right)\triangleq\begin{bmatrix}f^{T}\left(e+x_{d}\right)-h_{d}^{T}+u_{d}^{T}\left(x_{d}\right)g^{T}\left(e+x_{d}\right) & h_{d}^{T}\end{bmatrix}^{T}
\]
and 
\[
G\left(\zeta\right)\triangleq\begin{bmatrix}g^{T}\left(e+x_{d}\right) & \mathbf{0}_{m\times n}\end{bmatrix}^{T}.
\]
 The control error $\mu$ is treated hereafter as the design variable.
The control objective is to solve the infinite-horizon optimal regulation
problem online, i.e., to simultaneously synthesize and utilize a control
signal $\mu$ online to minimize the cost functional 
\[
J\left(\zeta,\mu\right)\triangleq\intop_{t_{0}}^{\infty}r\left(\zeta\left(\tau\right),\mu\left(\tau\right)\right)d\tau,
\]
 under the dynamic constraint 
\[
\dot{\zeta}=F\left(\zeta\right)+G\left(\zeta\right)\mu,
\]
 while tracking the desired trajectory, where $r:\mathbb{R}^{2n}\times\mathbb{R}^{m}\to\mathbb{R}$
is the local cost defined as 
\[
r\left(\zeta,\mu\right)\triangleq Q\left(e\right)+\mu^{T}R\mu,
\]
 $R\in\mathbb{R}^{m\times m}$ is a positive definite symmetric matrix
of constants, and $Q:\mathbb{R}^{n}\to\mathbb{R}$ is a continuous
positive definite function.

Assuming that an optimal policy exists, the optimal policy can be
characterized in terms of the value function $V^{*}:\mathbb{R}^{2n}\to\mathbb{R}$
defined as 
\[
V^{*}\!\left(\zeta\right)\!\triangleq\!\!\min_{\mu\left(\tau\right)\in U\mid\tau\in\mathbb{R}_{\geq t}}\intop_{t}^{\infty}\! r\left(\phi^{\mu}\!\left(\tau,t,\zeta\right),\mu\left(\tau\right)\right)d\tau,
\]
 where $U\in\mathbb{R}^{m}$ is the action space and the notation
$\phi^{\mu}\left(t;t_{0},\zeta_{0}\right)$ denotes the trajectory
of $\dot{\zeta}=F\left(\zeta\right)+G\left(\zeta\right)\mu,$ under
the control signal $\mu:\mathbb{R}_{\geq0}\to\mathbb{R}^{m}$ with
the initial condition $\zeta_{0}\in\mathbb{R}^{2n}$ and initial time
$t_{0}\in\mathbb{R}_{\geq0}$. Assuming that a minimizing policy exists
and that $V^{*}$ is continuously differentiable, a closed-form solution
for the optimal policy can be obtained as \cite{Kirk2004} 
\[
\mu^{*}\left(\zeta\right)=-\frac{1}{2}R^{-1}G^{T}\left(\zeta\right)\left(\nabla_{\zeta}V^{*}\left(\zeta\right)\right)^{T},
\]
 where $\nabla_{\zeta}\left(\cdot\right)\triangleq\frac{\partial\left(\cdot\right)}{\partial\zeta}$.
The optimal policy and the optimal value function satisfy the Hamilton-Jacobi-Bellman
(HJB) equation \cite{Kirk2004} 
\begin{equation}
\nabla\!_{\zeta}V^{*}\!\left(\zeta\right)\!\left(\! F\!\left(\zeta\right)\!+\! G\!\left(\zeta\right)\mu^{*}\!\left(\zeta\right)\!\right)\!+\overline{Q}\!\left(\zeta\right)\!+\!\mu^{*T}\!\left(\zeta\right)\! R\mu^{*}\!\left(\zeta\right)\!=\!0,\label{eq:CLTHJB}
\end{equation}
 with the initial condition $V^{*}\left(0\right)=0$, where the function
$\overline{Q}:\mathbb{R}^{2n}\to\mathbb{R}$ is defined as 
\[
\overline{Q}\left(\begin{bmatrix}e^{T} & x_{d}^{T}\end{bmatrix}^{T}\right)=Q\left(e\right),\:\forall e,\: x_{d}\in\mathbb{R}^{n}.
\]

\begin{rem}
Assumptions \ref{ass:CLTg} and \ref{ass:CLTAhd} can be eliminated
if a discounted cost optimal tracking problem is considered instead
of the total cost problem considered in this article. The discounted
cost tracking problem considers a value function of the form 
\[
V^{*}\!\left(\zeta\right)\!\triangleq\!\!\min_{u\left(\tau\right)\in U\mid\tau\in\mathbb{R}_{\geq t}}\intop_{t}^{\infty}\! e^{\kappa(t-\tau)}r\left(\phi^{u}\!\left(\tau,t,\zeta\right),u\left(\tau\right)\right)d\tau,
\]
 where $\kappa\in\mathbb{R}_{>0}$ is a constant discount factor,
and the control effort $u$ is minimized instead of the control error
$\mu.$ The control effort required for a system to perfectly track
a desired trajectory is generally nonzero even if the initial system
state is on the desired trajectory. Hence, in general, the optimal
value function for a discounted cost problem does not satisfy $V^{*}\left(0\right)=0$.
Online continuous-time RL techniques are generally analyzed using
the optimal value function as a candidate Lyapunov function. Since
the optimal value function for a discounted cost problem does not
evaluate to zero at the origin, it can not be used as a candidate
Lyapunov function. Hence, analyzing the stability of a discounted
cost optimal controller during the learning phase is complex.
\end{rem}

\section{Bellman Error}

Since a closed-form solution of the HJB is generally infeasible to
obtain, an approximate solution is sought. In an approximate actor-critic-based
solution, the optimal value function $V^{*}$ is replaced by a parametric
estimate $\hat{V}\left(\zeta,\hat{W}_{c}\right)$ and the optimal
policy $\mu^{*}$ by a parametric estimate $\hat{\mu}\left(\zeta,\hat{W}_{a}\right)$,
where $\hat{W}_{c}\in\mathbb{R}^{L}$ and $\hat{W}_{a}\in\mathbb{R}^{L}$
denote vectors of estimates of the ideal parameters. The objective
of the critic is to learn the parameters $\hat{W}_{c}$, and the objective
of the actor is to learn the parameters $\hat{W}_{a}$. Substituting
the estimates $\hat{V}$ and $\hat{\mu}$ for $V^{*}$ and $\mu^{*}$
in the HJB equation, respectively, yields a residual error $\delta:\mathbb{R}^{2n}\times\mathbb{R}^{L}\times\mathbb{R}^{L}\to\mathbb{R}$,
called the BE, is defined as 
\begin{multline}
\delta\left(\zeta,\hat{W}_{c},\hat{W}_{a}\right)=\overline{Q}\left(\zeta\right)+\hat{\mu}^{T}\left(\zeta,\hat{W}_{a}\right)R\hat{\mu}\left(\zeta,\hat{W}_{a}\right)\\
+\nabla_{\zeta}\hat{V}\left(\zeta,\hat{W}_{c}\right)\left(F\left(\zeta\right)+G\left(\zeta\right)\hat{\mu}\left(\zeta,\hat{W}_{a}\right)\right).\label{eq:CLTdelta}
\end{multline}
Specifically, to solve the optimal control problem, the critic aims
to find a set of parameters $\hat{W}_{c}$ and the actor aims to find
a set of parameters $\hat{W}_{a}$ such that 
\[
\delta\!\left(\!\zeta,\!\hat{W}_{c},\!\hat{W}_{a}\!\right)\!=\!0,
\]
 and 
\[
\hat{u}\left(\zeta,\hat{W}_{a}\right)\!=\!-\frac{1}{2}R^{-1}G^{T}\!\left(\zeta\right)\!\left(\nabla_{\zeta}\hat{V}\!\left(\!\zeta,\hat{W}_{a}\right)\!\right)^{T},
\]
 for all $\zeta\in\mathbb{R}^{2n}$. Since an exact basis for value
function approximation is generally not available, an approximate
set of parameters that minimizes the BE is sought. In particular,
to ensure uniform approximation of the value function and the policy
over a compact operating domain $\mathcal{C}\subset\mathbb{R}^{2n}$,
it is desirable to find parameters that minimize the error $E_{s}:\mathbb{R}^{L}\times\mathbb{R}^{L}\to\mathbb{R}$
defined as 
\[
E_{s}\left(\hat{W}_{c},\hat{W}_{a}\right)\triangleq\sup_{\zeta\in\mathcal{C}}\left|\delta\left(\zeta,\hat{W}_{c},\hat{W}_{a}\right)\right|.
\]
 Computation of the error $E_{s}$, and computation of the control
signal $u$ require knowledge of the system drift dynamics $f$. Two
prevalent approaches employed to render the control design robust
to uncertainties in the system drift dynamics are integral RL (cf.
\cite{Modares.Lewis.ea2014} and \cite{Lewis.Vrabie.ea2012}) and
state derivative estimation (cf. \cite{Bhasin.Kamalapurkar.ea2013a}
and \cite{Kamalapurkar.Dinh.ea2015}). 

Integral RL exploits the fact that for all $T>0$ and $t>t_{0}+T$,
the BE in (\ref{eq:CLTdelta}) has an equivalent integral form 
\begin{multline*}
\delta_{int}\left(t,\hat{W}_{c},\hat{W}_{a}\right)=\hat{V}\left(\phi^{\hat{\mu}}\left(t-T,t_{0},\zeta_{0}\right),\hat{W}_{c}\right)\\
-\intop_{t-T}^{t}r\left(\phi^{\hat{\mu}}\left(\tau,t_{0},\zeta_{0}\right),\hat{\mu}\left(\phi^{\hat{\mu}}\left(\tau,t_{0},\zeta_{0}\right),\hat{W}_{a}\right)\right)d\tau\\
-\hat{V}\left(\phi^{\hat{\mu}}\left(t,t_{0},\zeta_{0}\right),\hat{W}_{c}\right).
\end{multline*}
 Since the integral form does not require model knowledge, policies
designed based on $\delta_{int}$ can be implemented without knowledge
of $f.$

State derivative estimation-based techniques exploit the fact that
the BE in (\ref{eq:CLTdelta}) can be expressed as 
\begin{multline*}
\delta_{d}\left(\zeta,\dot{\zeta},\hat{W}_{a},\hat{W}_{c}\right)=\nabla_{\zeta}\hat{V}\left(\zeta,\hat{W}_{c}\right)\dot{\zeta}+\overline{Q}\left(\zeta\right)\\
+\hat{\mu}^{T}\left(\zeta,\hat{W}_{a}\right)R\hat{\mu}\left(\zeta,\hat{W}_{a}\right).
\end{multline*}
Hence, an estimate of the BE can be computed without model knowledge
if an estimate of the derivative $\dot{\zeta}$ is available. An adaptive
derivative estimator such as \cite{Bhasin.Kamalapurkar.ea2013b} could
be used to estimate $\dot{\zeta}$ online. 

The integral form of the BE is inherently dependent on the state trajectory,
and since adaptive derivative estimators approximate the derivative
only along the trajectory, derivative estimation-based techniques
are also dependent on the state trajectory. Hence, in techniques such
as \cite{Lewis.Vrabie.ea2012,Bhasin.Kamalapurkar.ea2013a,Modares.Lewis.ea2014,Kamalapurkar.Dinh.ea2015}
the BE can only be evaluated along the system trajectory. Thus, the
error $E_{s}$ is approximated by the instantaneous integral error
\[
\hat{E}\left(t\right)\triangleq\intop_{t_{0}}^{t}\delta^{2}\left(\phi^{\hat{\mu}}\left(\tau,t_{0},\zeta_{0}\right),\hat{W}_{c}\left(t\right),\hat{W}_{a}\left(t\right)\right)d\tau.
\]

Intuitively, for $\hat{E}$ to approximate $E$ over an operating
domain, the state trajectory $\phi^{\hat{\mu}}\left(t,t_{0},\zeta_{0}\right)$
needs to visit as many points in the operating domain as possible.
This intuition is formalized by the fact that techniques such as \cite{Lewis.Vrabie.ea2012,Bhasin.Kamalapurkar.ea2013a,Modares.Lewis.ea2014,Modares.Lewis2014,Kamalapurkar.Dinh.ea2015}
require PE to achieve convergence. The PE condition is relaxed in
\cite{Modares.Lewis.ea2014} to a finite excitation condition by using
integral RL along with experience replay, where each evaluation of
the BE $\delta_{int}$ is interpreted as gained experience, and these
experiences are stored in a history stack and are repeatedly used
in the learning algorithm to improve data efficiency. 

In this paper, a different approach is used to improve data efficiency.
A dynamic system identifier is developed to generate a parametric
estimate $\hat{F}\left(\zeta,\hat{\theta}\right)$ of the drift dynamics
$F$, where $\hat{\theta}$ denotes the estimate of the matrix of
unknown parameters. Given $\hat{F},$ $\hat{V}$, and $\hat{\mu}$,
an estimate of the BE can be evaluated at any $\zeta\in\mathbb{R}^{2n}$.
That is, using $\hat{F}$, experience can be simulated by extrapolating
the BE over unexplored off-trajectory points in the operating domain.
Hence, if an identifier can be developed such that $\hat{F}$ approaches
$F$ exponentially fast, learning laws for the optimal policy can
utilize simulated experience along with experience gained and stored
along the state trajectory. 

If parametric approximators are used to approximate $F$, convergence
of $\hat{F}$ to $F$ is implied by convergence of the parameters
to their unknown ideal values. It is well known that adaptive system
identifiers require PE to achieve parameter convergence. To relax
the PE condition, a CL-based (cf.\cite{Chowdhary2010a,Chowdhary.Johnson2011a,Chowdhary.Yucelen.ea2012,Kamalapurkar.Andrews.ea2014})
system identifier that uses recorded data for learning is developed
in the following section.

\section{System Identification\label{sec:CLTSystem-Identification}}

On any compact set $\mathcal{C}\subset\mathbb{R}^{n}$ the function
$f$ can be represented using a neural network (NN) as 
\[
f\left(x\right)=\theta^{T}\sigma_{f}\left(Y^{T}x_{1}\right)+\epsilon_{\theta}\left(x\right),
\]
 where $x_{1}\triangleq\begin{bmatrix}1 & x^{T}\end{bmatrix}^{T}\in\mathbb{R}^{n+1}$,
$\theta\in\mathbb{R}^{p+1\times n}$ and $Y\in\mathbb{R}^{n+1\times p}$
denote the constant unknown output-layer and hidden-layer NN weights,
$\sigma_{f}:\mathbb{R}^{p}\to\mathbb{R}^{p+1}$ denotes a bounded
NN basis function, $\epsilon_{\theta}:\mathbb{R}^{n}\to\mathbb{R}^{n}$
denotes the function reconstruction error, and $p\in\mathbb{N}$ denotes
the number of NN neurons. Using the universal function approximation
property of single layer NNs, given a constant matrix $Y$ such that
the rows of $\sigma_{f}\left(Y^{T}x_{1}\right)$ form a proper basis,
there exist constant ideal weights $\theta$ and known constants $\overline{\theta}$,
$\overline{\epsilon_{\theta}}$, and $\overline{\epsilon_{\theta}^{\prime}}\in\mathbb{R}$
such that $\left\Vert \theta\right\Vert _{F}\leq\overline{\theta}<\infty$,
$\sup_{x\in\mathcal{C}}\left\Vert \epsilon_{\theta}\left(x\right)\right\Vert \leq\overline{\epsilon_{\theta}}$,
and $\sup_{x\in\mathcal{C}}\left\Vert \nabla_{x}\epsilon_{\theta}\left(x\right)\right\Vert \leq\overline{\epsilon_{\theta}^{\prime}}$,
where $\left\Vert \cdot\right\Vert _{F}$ denotes the Frobenius norm
\cite{Lewis1999a}.

Using an estimate $\hat{\theta}\in\mathbb{R}^{p+1\times n}$ of the
weight matrix $\theta,$ the function $f$ can be approximated by
the function $\hat{f}:\mathbb{R}^{2n}\times\mathbb{R}^{p+1\times n}\to\mathbb{R}^{n}$
defined as 
\begin{equation}
\hat{f}\left(\zeta,\hat{\theta}\right)\triangleq\hat{\theta}^{T}\sigma_{\theta}\left(\zeta\right),\label{eq:CLTfhatnn}
\end{equation}
 where $\sigma_{\theta}:\mathbb{R}^{2n}\to\mathbb{R}^{p+1}$ is defined
as $\sigma_{\theta}\left(\zeta\right)=\sigma_{f}\left(Y^{T}\left[\begin{array}{cc}
1 & e^{T}+x_{d}^{T}\end{array}\right]^{T}\right)$. An estimator for online identification of the drift dynamics is
developed as 
\begin{equation}
\dot{\hat{x}}=\hat{\theta}^{T}\sigma_{\theta}\left(\zeta\right)+g\left(x\right)u+k\tilde{x},\label{eq:CLTsysid}
\end{equation}
where $\tilde{x}\triangleq x-\hat{x}$, and $k\in\mathbb{R}$ is a
positive constant learning gain. 
\begin{assumption}
\cite{Chowdhary.Johnson2011a} \label{ass:CLTsigmabar}A history stack
containing recorded state-action pairs $\left\{ x_{j},u_{j}\right\} _{j=1}^{M}$
along with numerically computed state derivatives $\left\{ \dot{\bar{x}}_{j}\right\} _{j=1}^{M}$
that satisfies 
\[
\lambda_{\min}\left(\sum_{j=1}^{M}\sigma_{fj}\sigma_{fj}^{T}\right)=\underline{\sigma_{\theta}}>0,\quad\left\Vert \dot{\bar{x}}_{j}-\dot{x}_{j}\right\Vert <\overline{d},\:\forall j,
\]
is available a priori, where $\sigma_{fj}\triangleq\sigma_{f}\left(Y^{T}\begin{bmatrix}1 & x_{j}^{T}\end{bmatrix}^{T}\right)$,
$\overline{d}\in\mathbb{R}$ is a known positive constant, $\dot{x}_{j}=f\left(x_{j}\right)+g\left(x_{j}\right)u_{j}$,
and $\lambda_{\min}\left(\cdot\right)$ denotes the minimum eigenvalue.%
\footnote{A priori availability of the history stack is used for ease of exposition,
and is not necessary. Provided the system states are exciting over
a finite time interval $t\in\left[t_{0},t_{0}+\overline{t}\right]$
(versus $t\in\left[t_{0},\infty\right)$ as in traditional PE-based
approaches) the history stack can also be recorded online. The controller
developed in \cite{Kamalapurkar.Dinh.ea2015} can be used over the
time interval $t\in\left[t_{0},t_{0}+\overline{t}\right]$ while the
history stack is being recorded, and the controller developed in this
result can be used thereafter. The use of two different controllers
results in a switched system with one switching event. Since there
is only one switching event, the stability of the switched system
follows from the stability of the individual subsystems.%
}
\end{assumption}
The weight estimates $\hat{\theta}$ are updated using the following
CL-based update law: 
\begin{equation}
\dot{\hat{\theta}}\!=\!\Gamma_{\theta}\sigma_{f}\!\left(Y^{T}\! x_{1}\!\right)\!\tilde{x}^{T}\!\!+\! k_{\theta}\Gamma_{\theta}\!\sum_{j=1}^{M}\!\sigma_{fj}\!\left(\!\dot{\bar{x}}_{j}\!-\! g_{j}u_{j}\!-\!\hat{\theta}^{T}\!\sigma_{fj}\!\right)^{T}\!,\label{eq:CLTThetahatdot}
\end{equation}
where $k_{\theta}\in\mathbb{R}$ is a constant positive CL gain, and
$\Gamma_{\theta}\in\mathbb{R}^{p+1\times p+1}$ is a constant, diagonal,
and positive definite adaptation gain matrix. Using (\ref{eq:CLTfhatnn}),
the BE in (\ref{eq:CLTdelta}) can be approximated as
\begin{multline}
\hat{\delta}\left(\zeta,\hat{\theta},\hat{W}_{c},\hat{W}_{a}\right)=\overline{Q}\left(\zeta\right)+\hat{\mu}^{T}\left(\zeta,\hat{W}_{a}\right)R\hat{\mu}\left(\zeta,\hat{W}_{a}\right),\\
+\nabla_{\zeta}\hat{V}\left(\zeta,\hat{W}_{a}\right)\left(F_{\theta}\left(\zeta,\hat{\theta}\right)+F_{1}\left(\zeta\right)+G\left(\zeta\right)\hat{\mu}\left(\zeta,\hat{W}_{a}\right)\right)\label{eq:CLTdeltahat}
\end{multline}
where 
\[
F_{\theta}\left(\zeta,\hat{\theta}\right)\triangleq\left[\begin{gathered}\hat{\theta}^{T}\sigma_{\theta}\left(\zeta\right)-g\left(x\right)g^{+}\left(x_{d}\right)\hat{\theta}^{T}\sigma_{\theta}\left(\begin{bmatrix}\mathbf{0}_{n\times1}\\
x_{d}
\end{bmatrix}\right)\\
0
\end{gathered}
\right],
\]
and 
\[
F_{1}\left(\zeta\right)\triangleq\left[\begin{gathered}-h_{d}+g\left(e+x_{d}\right)g^{+}\left(x_{d}\right)h_{d}\\
h_{d}
\end{gathered}
\right].
\]

\section{Value function approximation}

Since $V^{*}$ and $\mu^{*}$ are functions of the state $\zeta,$
the minimization problem stated in Section \ref{sec:Problem-formulation-and}
is intractable. To obtain a finite-dimensional minimization problem,
the optimal value function is represented over any compact operating
domain $\mathcal{C}\subset\mathbb{R}^{2n}$ using a NN as $V^{*}\left(\zeta\right)=W^{T}\sigma\left(\zeta\right)+\epsilon\left(\zeta\right),$
where $W\in\mathbb{R}^{L}$ denotes a vector of unknown NN weights,
$\sigma:\mathbb{R}^{2n}\to\mathbb{R}^{L}$ denotes a bounded NN basis
function, $\epsilon:\mathbb{R}^{2n}\to\mathbb{R}$ denotes the function
reconstruction error, and $L\in\mathbb{N}$ denotes the number of
NN neurons. Using the universal function approximation property of
single layer NNs, for any compact set $\mathcal{C}\subset\mathbb{R}^{2n}$,
there exist constant ideal weights $W$ and known positive constants
$\overline{W}$, $\overline{\epsilon}$, and $\overline{\epsilon^{\prime}}\in\mathbb{R}$
such that $\left\Vert W\right\Vert \leq\overline{W}<\infty$, $\sup_{\zeta\in\mathcal{C}}\left\Vert \epsilon\left(\zeta\right)\right\Vert \leq\overline{\epsilon}$,
and $\sup_{\zeta\in\mathcal{C}}\left\Vert \nabla_{\zeta}\epsilon\left(\zeta\right)\right\Vert \leq\overline{\epsilon^{\prime}}$
\cite{Lewis1999a}.

A NN representation of the optimal policy is obtained as 
\begin{equation}
\mu^{*}\left(\zeta\right)=-\frac{1}{2}R^{-1}G^{T}\left(\zeta\right)\left(\nabla_{\zeta}\sigma^{T}\left(\zeta\right)W+\nabla_{\zeta}\epsilon^{T}\left(\zeta\right)\right).\label{eq:CLTmu*nn}
\end{equation}
Using estimates $\hat{W}_{c}$ and $\hat{W}_{a}$ for the ideal weights
$W$, the optimal value function and the optimal policy are approximated
as
\begin{align}
\hat{V}\left(\zeta,\hat{W}_{c}\right) & \triangleq\hat{W}_{c}^{T}\sigma\left(\zeta\right),\nonumber \\
\hat{\mu}\left(\zeta,\hat{W}_{a}\right) & \triangleq-\frac{1}{2}R^{-1}G^{T}\left(\zeta\right)\nabla_{\zeta}\sigma^{T}\left(\zeta\right)\hat{W}_{a}.\label{eq:CLTVhatmuhat}
\end{align}
The optimal control problem is thus reformulated as the need to find
a set of weights $\hat{W}_{c}$ and $\hat{W}_{a}$ online, to minimize
the error 
\[
\hat{E}_{\hat{\theta}}\left(\hat{W}_{c},\hat{W}_{a}\right)\triangleq\sup_{\zeta\in\chi}\left|\hat{\delta}\left(\zeta,\hat{\theta},\hat{W}_{c},\hat{W}_{a}\right)\right|,
\]
 for a given $\hat{\theta}$, while simultaneously improving $\hat{\theta}$
using (\ref{eq:CLTThetahatdot}), and ensuring stability of the system
using the control law 
\begin{equation}
u=\hat{\mu}\left(\zeta,\hat{W}_{a}\right)+\hat{u}_{d}\left(\zeta,\hat{\theta}\right),\label{eq:CLTcontrollaw}
\end{equation}
where 
\[
\hat{u}_{d}\left(\zeta,\hat{\theta}\right)\triangleq g_{d}^{+}\left(h_{d}-\hat{\theta}^{T}\sigma_{\theta d}\right),
\]
 and $\sigma_{\theta d}\triangleq\sigma_{\theta}\left(\begin{bmatrix}\mathbf{0}_{1\times n} & x_{d}^{T}\end{bmatrix}^{T}\right)$.
The error between $u_{d}$ and $\hat{u}_{d}$ is included in the stability
analysis based on the fact that the error trajectories generated by
the system $\dot{e}=f\left(x\right)+g\left(x\right)u-\dot{x}_{d}$
under the controller in (\ref{eq:CLTcontrollaw}) are identical to
the error trajectories generated by the system $\dot{\zeta}=F\left(\zeta\right)+G\left(\zeta\right)\mu$
under the control law 
\begin{equation}
\mu=\hat{\mu}\left(\zeta,\hat{W}_{a}\right)+g_{d}^{+}\tilde{\theta}^{T}\sigma_{\theta d}+g_{d}^{+}\epsilon_{\theta d},\label{eq:CLTmuimpl}
\end{equation}
 where $\epsilon_{\theta d}\triangleq\epsilon_{\theta}\left(x_{d}\right)$.

\section{Simulation of experience}

Since computation of the supremum in $\hat{E}_{\hat{\theta}}$ is
intractable in general, simulation of experience is implemented by
minimizing a squared sum of BEs over finitely many points in the state
space. The following assumption facilitates the aforementioned approximation.
\begin{assumption}
\label{ass:CLTcbar}\cite{Kamalapurkar.Andrews.ea2014} There exists
a finite set of points $\left\{ \zeta_{i}\in\mathcal{C}\mid i=1,\cdots,N\right\} $
and a constant $\underline{c}\in\mathbb{R}$ such that 
\[
0<\underline{c}\triangleq\frac{1}{N}\left(\inf_{t\in\mathbb{R}_{\geq t_{0}}}\left(\lambda_{min}\left\{ \sum_{i=1}^{N}\frac{\omega_{i}\omega_{i}^{T}}{\rho_{i}}\right\} \right)\right),
\]
 where $\rho_{i}\triangleq1+\nu\omega_{i}^{T}\Gamma\omega_{i}\in\mathbb{R}$,
and 
\[
\omega_{i}\triangleq\nabla_{\zeta}\sigma\left(\zeta_{i}\right)\left(F_{\theta}\left(\zeta_{i},\hat{\theta}\right)+F_{1}\left(\zeta_{i}\right)+G\left(\zeta_{i}\right)\hat{\mu}\left(\zeta_{i},\hat{W}_{a}\right)\right).
\]
 
\end{assumption}
Using Assumption \ref{ass:CLTcbar}, simulation of experience is implemented
by the weight update laws 
\begin{align}
\dot{\hat{W}}_{c} & =-\eta_{c1}\Gamma\frac{\omega}{\rho}\hat{\delta}_{t}-\frac{\eta_{c2}}{N}\Gamma\sum_{i=1}^{N}\frac{\omega_{i}}{\rho_{i}}\hat{\delta}_{ti},\label{eq:CLTWchatdot}\\
\dot{\Gamma} & =\left(\beta\Gamma-\eta_{c1}\Gamma\frac{\omega\omega^{T}}{\rho^{2}}\Gamma\right)\mathbf{1}_{\left\{ \left\Vert \Gamma\right\Vert \leq\overline{\Gamma}\right\} },\:\left\Vert \Gamma\left(t_{0}\right)\right\Vert \leq\overline{\Gamma},\label{eq:CLTGammadot}\\
\dot{\hat{W}}_{a} & =-\eta_{a1}\left(\hat{W}_{a}-\hat{W}_{c}\right)-\eta_{a2}\hat{W}_{a}\nonumber \\
 & +\left(\frac{\eta_{c1}G_{\sigma}^{T}\hat{W}_{a}\omega^{T}}{4\rho}+\sum_{i=1}^{N}\frac{\eta_{c2}G_{\sigma i}^{T}\hat{W}_{a}\omega_{i}^{T}}{4N\rho_{i}}\right)\hat{W}_{c},\label{eq:CLTWahatdot}
\end{align}
where 
\[
\omega\triangleq\nabla_{\zeta}\sigma\left(\zeta\right)\left(F_{\theta}\left(\zeta,\hat{\theta}\right)+F_{1}\left(\zeta\right)+G\left(\zeta\right)\hat{\mu}\left(\zeta,\hat{W}_{a}\right)\right),
\]
$\Gamma\in\mathbb{R}^{L\times L}$ is the least-squares gain matrix,
$\overline{\Gamma}\in\mathbb{R}$ denotes a positive saturation constant,
$\beta\in\mathbb{R}$ denotes a constant forgetting factor, $\eta_{c1},\eta_{c2},\eta_{a1},\eta_{a2}\in\mathbb{R}$
denote constant positive adaptation gains, $\mathbf{1}_{\left\{ \cdot\right\} }$
denotes the indicator function of the set $\left\{ \cdot\right\} $,
$G_{\sigma}\triangleq\nabla_{\zeta}\mbox{\ensuremath{\sigma}}\left(\zeta\right)G\left(\zeta\right)R^{-1}G^{T}\left(\zeta\right)\nabla_{\zeta}\sigma^{T}\left(\zeta\right)$,
and $\rho\triangleq1+\nu\omega^{T}\Gamma\omega$, where $\nu\in\mathbb{R}$
is a positive normalization constant. In (\ref{eq:CLTWchatdot})-(\ref{eq:CLTWahatdot})
and in the subsequent development, for any function $\xi\left(\zeta,\cdot\right)$,
the notation $\xi_{i}$, is defined as $\xi_{i}\triangleq\xi\left(\zeta_{i},\cdot\right)$,
and the instantaneous BEs $\hat{\delta}_{t}$ and $\hat{\delta}_{ti}$
are given by 
\[
\hat{\delta}_{t}=\hat{\delta}\left(\zeta,\hat{W}_{c},\hat{W}_{a},\hat{\theta}\right),
\]
 and 
\[
\hat{\delta}_{ti}=\hat{\delta}\left(\zeta_{i},\hat{W}_{c},\hat{W}_{a},\hat{\theta}\right).
\]

\section{Stability analysis}

If the state penalty function $\overline{Q}$ is positive definite,
then the optimal value function $V^{*}$ is positive definite, and
serves as a Lyapunov function for the concatenated system under the
optimal control policy $\mu^{*}$; hence, $V^{*}$ is used (cf. \cite{Vamvoudakis2010,Lewis.Vrabie.ea2012,Bhasin.Kamalapurkar.ea2013a})
as a candidate Lyapunov function for the closed-loop system under
the policy $\hat{\mu}.$ The function $\overline{Q}$, and hence,
the function $V^{*}$ are positive semidefinite; hence, the function
$V^{*}$ is not a valid candidate Lyapunov function. However, the
results in \cite{Kamalapurkar.Dinh.ea2015} can be used to show that
a nonautonomous form of the optimal value function denoted by $V_{t}^{*}:\mathbb{R}^{n}\times\mathbb{R}\to\mathbb{R}$,
defined as 
\[
V_{t}^{*}\left(e,t\right)=V^{*}\left(\begin{bmatrix}e\\
x_{d}\left(t\right)
\end{bmatrix}\right),\:\forall e\in\mathbb{R}^{n},\: t\in\mathbb{R},
\]
is positive definite and decrescent. Hence, $V_{t}^{*}\left(0,t\right)=0,\:\forall t\in\mathbb{R}$
and there exist class $\mathcal{K}$ functions $\underline{v}:\mathbb{R}\to\mathbb{R}$
and $\overline{v}:\mathbb{R}\to\mathbb{R}$ such that 
\begin{equation}
\underline{v}\left(\left\Vert e\right\Vert \right)\leq V_{t}^{*}\left(e,t\right)\le\overline{v}\left(\left\Vert e\right\Vert \right),\label{eq:CLTVt*bounds}
\end{equation}
for all $e\in\mathbb{R}^{n}$ and for all $t\in\mathbb{R}$.

To facilitate the stability analysis, a candidate Lyapunov function
$V_{0}:\mathbb{R}^{n}\times\mathbb{R}^{p+1\times n}\to\mathbb{R}$
is selected as
\begin{equation}
V_{0}\left(\tilde{x},\tilde{\theta}\right)=\frac{1}{2}\tilde{x}^{T}\tilde{x}+\frac{1}{2}\mbox{tr}\left(\tilde{\theta}^{T}\Gamma_{\theta}^{-1}\tilde{\theta}\right),\label{eq:CLTV0}
\end{equation}
where $\tilde{\theta}\triangleq\theta-\hat{\theta}$ and $\mbox{tr}\left(\cdot\right)$
denotes the trace of a matrix. Using (\ref{eq:CLTsysid})-(\ref{eq:CLTThetahatdot}),
the following bound on the time derivative of $V_{0}$ is established:
\begin{gather}
\dot{V}_{0}\leq-k\left\Vert \tilde{x}\right\Vert ^{2}-k_{\theta}\underline{\sigma_{\theta}}\left\Vert \tilde{\theta}\right\Vert _{F}^{2}+\overline{\epsilon_{\theta}}\left\Vert \tilde{x}\right\Vert +k_{\theta}\overline{d_{\theta}}\left\Vert \tilde{\theta}\right\Vert _{F},\label{eq:CLTV0Dot}
\end{gather}
where 
\[
\overline{d_{\theta}}\triangleq\overline{d}\sum_{j=1}^{M}\left\Vert \sigma_{\theta}{}_{j}\right\Vert +\sum_{j=1}^{M}\left(\left\Vert \epsilon_{\theta}{}_{j}\right\Vert \left\Vert \sigma_{\theta}{}_{j}\right\Vert \right).
\]

A concatenated state $Z\in\mathbb{R}^{2n+2L+n\left(p+1\right)}$ is
defined as 
\[
Z\triangleq\begin{bmatrix}e^{T} & \tilde{W}_{c}^{T} & \tilde{W}_{a}^{T} & \tilde{x}^{T} & \left(\mbox{vec}\left(\tilde{\theta}\right)\right)^{T}\end{bmatrix}^{T},
\]
and a candidate Lyapunov function is defined as 
\begin{equation}
V_{L}\!\left(Z,t\right)\!\triangleq\! V_{t}^{*}\!\left(e,t\right)+\frac{1}{2}\tilde{W}_{c}^{T}\Gamma^{-1}\tilde{W}_{c}+\frac{1}{2}\tilde{W}_{a}^{T}\tilde{W}_{a}+V_{0}\!\left(\!\tilde{\theta},\tilde{x}\!\right)\!,\label{eq:CLTVL}
\end{equation}
where $\mbox{vec}\left(\cdot\right)$ denotes the vectorization operator
and $V_{0}$ is defined in (\ref{eq:CLTV0}). The saturated least-squares
update law in (\ref{eq:CLTGammadot}) ensures that there exist positive
constants $\underline{\gamma},\overline{\gamma}\in\mathbb{R}$ such
that
\begin{equation}
\underline{\gamma}\leq\left\Vert \Gamma^{-1}\left(t\right)\right\Vert \leq\overline{\gamma},\:\forall t\in\mathbb{R}.\label{eq:CLTGammabound}
\end{equation}
Using (\ref{eq:CLTV0}), the bounds in (\ref{eq:CLTGammabound}) and
(\ref{eq:CLTVt*bounds}), and the fact that $\mbox{tr}\left(\tilde{\theta}^{T}\Gamma_{\theta}^{-1}\tilde{\theta}\right)=\left(\mbox{vec}\left(\tilde{\theta}\right)\right)^{T}\left(\Gamma_{\theta}^{-1}\otimes\mathbb{I}_{p+1}\right)\left(\mbox{vec}\left(\tilde{\theta}\right)\right)$,
the candidate Lyapunov function in (\ref{eq:CLTVL}) can be bounded
as
\begin{equation}
\underline{v_{l}}\left(\left\Vert Z\right\Vert \right)\leq V_{L}\left(Z,t\right)\leq\overline{v_{l}}\left(\left\Vert Z\right\Vert \right),\label{eq:CLTVLbounds}
\end{equation}
for all $Z\in\mathbb{R}^{2n+2L+n\left(p+1\right)}$ and for all $t\in\mathbb{R}$,
where $\underline{v_{l}}:\mathbb{R}\to\mathbb{R}$ and $\overline{v_{l}}:\mathbb{R}\to\mathbb{R}$
are class $\mathcal{K}$ functions. 

For notational brevity, the dependence of the functions $F,$ $G,$
$\sigma,$ $\sigma^{\prime},$ $\epsilon,$ $\epsilon^{\prime},$
$\sigma_{\theta},$ $\epsilon_{\theta},$ and $g$ on the system states
is suppressed hereafter. To facilitate the stability analysis, the
approximate BE in (\ref{eq:CLTdeltahat}) is expressed in terms of
the weight estimation errors as
\begin{equation}
\hat{\delta}_{t}=-\omega^{T}\tilde{W}_{c}-W^{T}\sigma^{\prime}F_{\tilde{\theta}}+\frac{1}{4}\tilde{W}_{a}^{T}G_{\sigma}\tilde{W}_{a}+\Delta,\label{eq:CLTdeltahatunmeas}
\end{equation}
where $F_{\tilde{\theta}}\triangleq F_{\theta}\left(\zeta,\tilde{\theta}\right)$
and $\Delta=O\left(\overline{\epsilon},\overline{\epsilon^{\prime}},\overline{\epsilon_{\theta}}\right)$.
Given any compact set $\chi\subset\mathbb{R}^{2n+2L+n\left(p+1\right)}$
containing an open ball of radius $\rho\in\mathbb{R}$ centered at
the origin, a positive constant $\iota\in\mathbb{R}$ is defined as
\begin{multline}
\iota\triangleq\frac{3\left(\frac{\left(\eta_{c1}+\eta_{c2}\right)\overline{W}^{2}\overline{\left\Vert G_{\sigma}\right\Vert }}{16\sqrt{\nu\underline{\Gamma}}}+\frac{\overline{\left\Vert \left(W^{T}G_{\sigma}+\epsilon^{\prime}G_{r}\sigma^{\prime T}\right)\right\Vert }}{4}+\frac{\eta_{a2}\overline{W}}{2}\right)^{2}}{\left(\eta_{a1}+\eta_{a2}\right)}\\
+\frac{3\left(\left(\overline{\left\Vert W^{T}\sigma^{\prime}Gg_{d}^{+}\right\Vert }+\overline{\left\Vert \epsilon^{\prime}Gg_{d}^{+}\right\Vert }\right)\overline{\sigma_{g}}+k_{\theta}\overline{d_{\theta}}\right)^{2}}{4k_{\theta}\underline{\sigma_{\theta}}}\\
+\frac{\left(\eta_{c1}+\eta_{c2}\right)^{2}\overline{\left\Vert \Delta\right\Vert }^{2}}{4\nu\underline{\Gamma}\eta_{c2}\underline{c}}+\frac{\overline{\epsilon_{\theta}}^{2}}{2k}+\overline{\left\Vert \epsilon^{\prime}Gg_{d}^{+}\epsilon_{\theta d}\right\Vert }\\
+\overline{\left\Vert \frac{1}{2}G_{\epsilon}\right\Vert }+\overline{\left\Vert \frac{1}{2}W^{T}\sigma^{\prime}G_{r}\epsilon^{\prime T}\right\Vert }+\overline{\left\Vert W^{T}\sigma^{\prime}Gg_{d}^{+}\epsilon_{\theta d}\right\Vert },\label{eq:CLTiota}
\end{multline}
where $G_{r}\triangleq GR^{-1}G^{T},$ and $G_{\epsilon}\triangleq\epsilon^{\prime}G_{r}\left(\epsilon^{\prime}\right)^{T}$.
Let $v_{l}:\mathbb{R}\to\mathbb{R}$ be a class $\mathcal{K}$ function
such that
\begin{multline}
v_{l}\left(\left\Vert Z\right\Vert \right)\leq\frac{\underline{q}\left(\left\Vert e\right\Vert \right)}{2}+\frac{\eta_{c2}\underline{c}}{8}\left\Vert \tilde{W}_{c}\right\Vert ^{2}+\frac{\left(\eta_{a1}+\eta_{a2}\right)}{6}\left\Vert \tilde{W}_{a}\right\Vert ^{2}\\
+\frac{k}{4}\left\Vert \tilde{x}\right\Vert ^{2}+\frac{k_{\theta}\underline{\sigma_{\theta}}}{6}\left\Vert \mbox{vec}\left(\tilde{\theta}\right)\right\Vert ^{2}.\label{eq:CLTnul}
\end{multline}
The sufficient gain conditions used in the subsequent Theorem \ref{thm:CLTMain}
are 
\begin{align}
v_{l}^{-1}\left(\iota\right) & <\overline{v_{l}}^{-1}\left(\underline{v_{l}}\left(\rho\right)\right)\label{eq:CLTchicond}\\
\eta_{c2}\underline{c} & >\frac{3\left(\eta_{c2}+\eta_{c1}\right)^{2}\overline{W}^{2}\overline{\left\Vert \sigma^{\prime}\right\Vert }^{2}\overline{\sigma_{g}}^{2}}{4k_{\theta}\underline{\sigma_{\theta}}\nu\underline{\Gamma}}\label{eq:CLTetaccond}\\
\left(\eta_{a1}+\eta_{a2}\right) & >\frac{3\left(\eta_{c1}+\eta_{c2}\right)\overline{W}\overline{\left\Vert G_{\sigma}\right\Vert }}{8\sqrt{\nu\underline{\Gamma}}}\nonumber \\
 & +\frac{3}{\underline{c}\eta_{c2}}\left(\frac{\left(\eta_{c1}+\eta_{c2}\right)\overline{W}\overline{\left\Vert G_{\sigma}\right\Vert }}{8\sqrt{\nu\underline{\Gamma}}}+\eta_{a1}\right)^{2}.\label{eq:CLTetaacond}
\end{align}
In (\ref{eq:CLTiota})-(\ref{eq:CLTetaacond}), for any function $\varpi:\mathbb{R}^{l}\to\mathbb{R}$,
$l\in\mathbb{N}$, the notation $\overline{\left\Vert \varpi\right\Vert }$,
denotes $\sup_{y\in\chi\cap\mathbb{R}^{l}}\left\Vert \varpi\left(y\right)\right\Vert $,
and $\overline{\sigma_{g}}\triangleq\overline{\left\Vert \sigma_{\theta}\right\Vert }+\overline{\left\Vert gg_{d}^{+}\right\Vert }\overline{\left\Vert \sigma_{\theta d}\right\Vert }.$ 

The sufficient condition in (\ref{eq:CLTchicond}) requires the set
$\chi$ to be large enough based on the constant $\iota$. Since the
NN approximation errors depend on the compact set $\chi$, in general,
for a fixed number of NN neurons, the constant $\iota$ increases
with the size of the set $\chi$. However, for a fixed set $\chi,$
the constant $\iota$ can be reduced by reducing function reconstruction
errors, i.e., by increasing number of NN neurons, and by increasing
the learning gains provided $\underline{\sigma_{\theta}}$ is large
enough. Hence a sufficient number of NN neurons and extrapolation
points are required to satisfy the condition in (\ref{eq:CLTchicond}).
\begin{thm}
\label{thm:CLTMain}Provided Assumptions \ref{ass:CLTAhd}-\ref{ass:CLTcbar}
hold, and the control gains, number of NN neurons, and BE extrapolation
points are selected based on (\ref{eq:CLTchicond})-(\ref{eq:CLTetaacond}),
the controller in (\ref{eq:CLTcontrollaw}), along with the weight
update laws (\ref{eq:CLTWchatdot})-(\ref{eq:CLTWahatdot}), and the
identifier in (\ref{eq:CLTsysid}) along with the weight update law
(\ref{eq:CLTThetahatdot}) ensure that the system states remain bounded,
the tracking error is ultimately bounded, and that the control policy
$\hat{\mu}$ converges to a neighborhood around the optimal control
policy $\mu^{*}.$\end{thm}
\begin{IEEEproof}
Using (\ref{eq:CLTzetadyn}) and the fact that 
\[
\dot{V}_{t}^{*}\left(e\left(t\right),t\right)=\dot{V}^{*}\left(\zeta\left(t\right)\right),\:\forall t\in\mathbb{R},
\]
 the time-derivative of the candidate Lyapunov function in (\ref{eq:CLTVL})
is 
\begin{multline}
\dot{V}_{L}=\nabla_{\zeta}V^{*}\left(F+G\mu^{*}\right)-\tilde{W}_{c}^{T}\Gamma^{-1}\dot{\hat{W}}_{c}-\frac{1}{2}\tilde{W}_{c}^{T}\Gamma^{-1}\dot{\Gamma}\Gamma^{-1}\tilde{W}_{c}\\
-\tilde{W}_{a}^{T}\dot{\hat{W}}_{a}+\dot{V}_{0}+\nabla_{\zeta}V^{*}G\mu-\nabla_{\zeta}V^{*}G\mu^{*}.\label{eq:CLTVLdot1}
\end{multline}
Using (\ref{eq:CLTHJB}), (\ref{eq:CLTmu*nn}), (\ref{eq:CLTVhatmuhat}),
and (\ref{eq:CLTmuimpl}) the expression in (\ref{eq:CLTVLdot1})
is bounded as
\begin{multline}
\dot{V}_{L}\leq-\overline{Q}\left(\zeta\right)-\tilde{W}_{c}^{T}\Gamma^{-1}\dot{\hat{W}}_{c}-\frac{1}{2}\tilde{W}_{c}^{T}\Gamma^{-1}\dot{\Gamma}\Gamma^{-1}\tilde{W}_{c}-\tilde{W}_{a}^{T}\dot{\hat{W}}_{a}\\
+\dot{V}_{0}+\frac{1}{2}\left(W^{T}G_{\sigma}+\epsilon^{\prime}G_{r}\sigma^{\prime T}\right)\tilde{W}_{a}+W^{T}\sigma^{\prime}Gg_{d}^{+}\tilde{\theta}^{T}\sigma_{\theta d}\\
+\epsilon^{\prime}Gg_{d}^{+}\tilde{\theta}^{T}\sigma_{\theta d}+\frac{1}{2}G_{\epsilon}+\frac{1}{2}W^{T}\sigma^{\prime}G_{r}\epsilon^{\prime T}+W^{T}\sigma^{\prime}Gg_{d}^{+}\epsilon_{\theta d}\\
-\left(\mu^{*}\right)^{T}R\mu^{*}+\epsilon^{\prime}Gg_{d}^{+}\epsilon_{\theta d}.\label{eq:CLTVLdot2}
\end{multline}
Using the update laws in (\ref{eq:CLTWchatdot})-(\ref{eq:CLTWahatdot}),
the bound in (\ref{eq:CLTV0Dot}), and (\ref{eq:CLTdeltahatunmeas}),
the expression in (\ref{eq:CLTVLdot2}) is bounded as
\begin{multline*}
\dot{V}_{L}\leq-\overline{Q}\left(\zeta\right)-\sum_{i=1}^{N}\tilde{W}_{c}^{T}\frac{\eta_{c2}}{N}\frac{\omega_{i}\omega_{i}^{T}}{\rho_{i}}\tilde{W}_{c}-k_{\theta}\underline{\sigma_{\theta}}\left\Vert \tilde{\theta}\right\Vert _{F}^{2}\\
-\left(\eta_{a1}+\eta_{a2}\right)\tilde{W}_{a}^{T}\tilde{W}_{a}-k\left\Vert \tilde{x}\right\Vert ^{2}-\eta_{c1}\tilde{W}_{c}^{T}\frac{\omega}{\rho}W^{T}\sigma^{\prime}F_{\tilde{\theta}}\\
+\eta_{c1}\tilde{W}_{c}^{T}\frac{\omega}{\rho}\Delta+\eta_{a1}\tilde{W}_{a}^{T}\tilde{W}_{c}+\eta_{a2}\tilde{W}_{a}^{T}W\\
+\frac{1}{4}\eta_{c1}\tilde{W}_{c}^{T}\frac{\omega}{\rho}\tilde{W}_{a}^{T}G_{\sigma}\tilde{W}_{a}-\sum_{i=1}^{N}\tilde{W}_{c}^{T}\frac{\eta_{c2}}{N}\frac{\omega_{i}}{\rho_{i}}W^{T}\sigma_{i}^{\prime}F_{\tilde{\theta}i}\\
+\sum_{i=1}^{N}\frac{1}{4}\tilde{W}_{c}^{T}\frac{\eta_{c2}}{N}\frac{\omega_{i}}{\rho_{i}}\tilde{W}_{a}^{T}G_{\sigma i}\tilde{W}_{a}+\tilde{W}_{c}^{T}\frac{\eta_{c2}}{N}\sum_{i=1}^{N}\frac{\omega_{i}}{\rho_{i}}\Delta_{i}\\
-\tilde{W}_{a}^{T}\left(\frac{\eta_{c1}G_{\sigma}^{T}\hat{W}_{a}\omega^{T}}{4\rho}+\sum_{i=1}^{N}\frac{\eta_{c2}G_{\sigma i}^{T}\hat{W}_{a}\omega_{i}^{T}}{4N\rho_{i}}\right)\hat{W}_{c}\\
+\overline{\epsilon_{\theta}}\left\Vert \tilde{x}\right\Vert +k_{\theta}\overline{d_{\theta}}\left\Vert \tilde{\theta}\right\Vert _{F}+\frac{1}{2}\left(W^{T}G_{\sigma}+\epsilon^{\prime}G_{r}\sigma^{\prime T}\right)\tilde{W}_{a}\\
+W^{T}\sigma^{\prime}Gg_{d}^{+}\tilde{\theta}^{T}\sigma_{\theta d}+\epsilon^{\prime}Gg_{d}^{+}\tilde{\theta}^{T}\sigma_{\theta d}+\frac{1}{2}G_{\epsilon}\\
+\frac{1}{2}W^{T}\sigma^{\prime}G_{r}\epsilon^{\prime T}+W^{T}\sigma^{\prime}Gg_{d}^{+}\epsilon_{\theta d}+\epsilon^{\prime}Gg_{d}^{+}\epsilon_{\theta d}.
\end{multline*}
Segregation of terms, completion of squares, and the use of Young's
inequalities yields
\begin{multline}
\dot{V}_{L}\leq-\overline{Q}\left(\zeta\right)-\frac{\eta_{c2}\underline{c}}{4}\left\Vert \tilde{W}_{c}\right\Vert ^{2}-\frac{\left(\eta_{a1}+\eta_{a2}\right)}{3}\left\Vert \tilde{W}_{a}\right\Vert ^{2}\\
-\frac{k}{2}\left\Vert \tilde{x}\right\Vert ^{2}-\frac{k_{\theta}\underline{\sigma_{\theta}}}{3}\left\Vert \tilde{\theta}\right\Vert _{F}^{2}\\
-\left(\frac{\eta_{c2}\underline{c}}{4}-\frac{3\left(\eta_{c2}+\eta_{c1}\right)^{2}\overline{W}^{2}\overline{\left\Vert \sigma^{\prime}\right\Vert }^{2}\overline{\sigma_{g}}^{2}}{16k_{\theta}\underline{\sigma_{\theta}}\nu\underline{\Gamma}}\right)\left\Vert \tilde{W}_{c}\right\Vert ^{2}\\
-\left(\frac{\left(\eta_{a1}+\eta_{a2}\right)}{3}-\frac{\left(\eta_{c1}+\eta_{c2}\right)\overline{W}\overline{\left\Vert G_{\sigma}\right\Vert }}{8\sqrt{\nu\underline{\Gamma}}}\right)\left\Vert \tilde{W}_{a}\right\Vert ^{2}\\
+\frac{1}{\underline{c}\eta_{c2}}\left(\frac{\left(\eta_{c1}+\eta_{c2}\right)\overline{W}\overline{\left\Vert G_{\sigma}\right\Vert }}{8\sqrt{\nu\underline{\Gamma}}}+\eta_{a1}\right)^{2}\left\Vert \tilde{W}_{a}\right\Vert ^{2}\\
+\frac{3\left(\frac{\left(\eta_{c1}+\eta_{c2}\right)\overline{W}^{2}\overline{\left\Vert G_{\sigma}\right\Vert }}{16\sqrt{\nu\underline{\Gamma}}}+\frac{\overline{\left\Vert \left(W^{T}G_{\sigma}+\epsilon^{\prime}G_{r}\sigma^{\prime T}\right)\right\Vert }}{4}+\frac{\eta_{a2}\left\Vert W\right\Vert }{2}\right)^{2}}{\left(\eta_{a1}+\eta_{a2}\right)}\\
+\frac{3\left(\left(\overline{\left\Vert W^{T}\sigma^{\prime}Gg_{d}^{+}\right\Vert }+\overline{\left\Vert \epsilon^{\prime}Gg_{d}^{+}\right\Vert }\right)\overline{\sigma_{g}}+k_{\theta}\overline{d_{\theta}}\right)^{2}}{4k_{\theta}\underline{\sigma_{\theta}}}\\
+\frac{\left(\eta_{c1}+\eta_{c2}\right)^{2}\overline{\left\Vert \Delta\right\Vert }^{2}}{4\nu\underline{\Gamma}\eta_{c2}\underline{c}}+\frac{\overline{\epsilon_{\theta}}^{2}}{2k}+\overline{\left\Vert \frac{1}{2}G_{\epsilon}\right\Vert }+\overline{\left\Vert \frac{1}{2}W^{T}\sigma^{\prime}G_{r}\epsilon^{\prime T}\right\Vert }\\
+\overline{\left\Vert W^{T}\sigma^{\prime}Gg_{d}^{+}\epsilon_{\theta d}\right\Vert }+\overline{\left\Vert \epsilon^{\prime}Gg_{d}^{+}\epsilon_{\theta d}\right\Vert },\label{eq:CLTVLdot3}
\end{multline}
for all $Z\in\mathbb{\chi}$. Provided the sufficient conditions in
(\ref{eq:CLTetaccond})-(\ref{eq:CLTetaacond}) are satisfied, the
expression in (\ref{eq:CLTVLdot3}) yields 
\begin{equation}
\dot{V}_{L}\leq-v_{l}\left(\left\Vert Z\right\Vert \right),\:\forall\left\Vert Z\right\Vert \geq v_{l}^{-1}\left(\iota\right),\:\forall Z\in\chi.\label{eq:CLTVLdot4}
\end{equation}
Using (\ref{eq:CLTVLbounds}), (\ref{eq:CLTchicond}), and (\ref{eq:CLTVLdot4})
Theorem 4.18 in \cite{Khalil2002} can be invoked to conclude that
every trajectory $Z\left(t\right)$ satisfying $\left\Vert Z\left(t_{0}\right)\right\Vert \leq\overline{v_{l}}^{-1}\left(\underline{v_{l}}\left(\rho\right)\right)$,
is bounded for all $t\in\mathbb{R}$ and satisfies $\lim\sup_{t\to\infty}\left\Vert Z\left(t\right)\right\Vert \leq\underline{v_{l}}^{-1}\left(\overline{v_{l}}\left(v_{l}^{-1}\left(\iota\right)\right)\right).$
\end{IEEEproof}

\section{Conclusion}

A concurrent-learning based implementation of model-based RL is developed
to obtain an approximate online solution to infinite horizon optimal
tracking problems for nonlinear continuous-time control-affine systems.
The desired steady-state controller is used to facilitate the formulation
of a feasible optimal control problem, and the system state is augmented
with the desired trajectory to facilitate the formulation of a stationary
optimal control problem. A CL-based system identifier is developed
to remove the dependence of the desired steady-state controller on
the system drift dynamics, and to facilitate simulation of experience
via BE extrapolation. Simulation results are provided to demonstrate
the effectiveness of the developed technique.%

Similar to the PE condition in RL-based online optimal control literature,
Assumption \ref{ass:CLTcbar} can not, in general, be guaranteed a
priori. However, Assumption \ref{ass:CLTcbar} can be heuristically
met by oversampling, i.e., by selecting $N\gg L.$ Furthermore, unlike
PE, the satisfaction of Assumption \ref{ass:CLTcbar} can be monitored
online; hence, threshold-based algorithms can be employed to preserve
rank by selecting new points if the minimum singular value falls below
a certain threshold. Provided the minimum singular value does not
decrease during a switch, the trajectories of the resulting switched
system can be shown to be uniformly bounded using a common Lyapunov
function. Formulation of sufficient conditions for Assumption \ref{ass:CLTcbar}
that can be verified a priori is a topic for future research.

\bibliographystyle{IEEEtran}
\bibliography{ncr,master,encr}

\end{document}